\pdfoutput=1				
\documentclass[twocolumn,twoside,slac_two]{revtex4}
\usepackage{graphicx}
\usepackage{fancyhdr}
\usepackage{graphics}
\usepackage{epstopdf}
\usepackage{textpos}
\usepackage{float}
\newcommand{\GeV}{~GeV}
\newcommand{\TeV}{~TeV}
\newcommand{\fbinv}{$\mathrm{fb^{-1}}$}

\newcommand{\anaLumi}{1.1~\fbinv}

\newcommand{\limitZssm}{1940}
\newcommand{\limitZpsi}{1620}
\newcommand{\limitGlow}{1450}
\newcommand{\limitGhigh}{1780}

\newcommand{\cPZpr}{\mathrm{Z}^\prime}
\newcommand{\cPZ}{\mathrm{Z}}
\newcommand{\Pgmp}{\mathrm{\mu^{+}}}
\newcommand{\Pgmm}{\mathrm{\mu^{-}}}
\newcommand{\Pgm}{\mathrm{\mu}}
\newcommand{\Pe}{\mathrm{e}}
\newcommand{\Pp}{\mathrm{p}}
\newcommand{\cPgn}{\mathrm{\nu}}
\newcommand{\ttbar}{\mathrm{t\bar{t}}}
\newcommand{\PW}{\mathrm{W}}
\newcommand{\Pgt}{\mathrm{\tau}}
\newcommand{\GKK}{\ensuremath{\mathrm{G}_\mathrm{KK}}}
\newcommand{\tq}{\ensuremath{\mathrm{t}}}
\newcommand{\ZPSSM}{\ensuremath{\mathrm{Z}^\prime_\mathrm{SSM}}}
\newcommand{\ZPPSI}{\ensuremath{\mathrm{Z}^\prime_\psi}}

\begin{document}

%%%%%%%%%%%%%%%%%%%%%% WRITE THE TITLE HERE %%%%%%%%%%%%%%%%%%%
\title{\centering Search for High-Mass Resonances in the Dilepton Final State with the CMS Detector.}
%%%%%%%%%%%%%%%%%%%%%% WRITE THE AUTHOR HERE %%%%%%%%%%%%%%%%%

%%% Please insert your personal picture here!

\author{
%\centering
%\includegraphics[scale=0.15]{mypicture.jpg} \\
\begin{center}
Vladlen Timciuc for the CMS Collaboration
\end{center}
}
\affiliation{
\centering
California Institute of Technology, Pasadena CA 91125 USA
}
%%%%%%%%%%%%%%%%%%%%%% WRITE THE ABSTRACT HERE %%%%%%%%%%%%%%%%
\begin{abstract}

   A search for narrow resonances at high mass in the dimuon and
  dielectron channels has been performed by the CMS experiment at the
  CERN LHC, using $\mathrm{pp}$ collision data recorded at
  $\sqrt{s}=7$~TeV. The event samples correspond to an integrated
  luminosity of \anaLumi.
  Heavy dilepton resonances are predicted
  in theoretical models with extra gauge bosons ($\mathrm{Z}^\prime$)
  or as Kaluza--Klein graviton excitations ($\mathrm{G}_\mathrm{KK}$)
  in the Randall-Sundrum model. Upper limits on the inclusive cross
  section of
  $\mathrm{Z}^\prime(\mathrm{G}_\mathrm{KK})\rightarrow\ell^+\ell^-$
  relative to $\mathrm{Z}\rightarrow\ell^+\ell^-$ are presented.
  These limits exclude at 95\% confidence level a $\mathrm{Z}^\prime$
  with standard-model-like couplings below \limitZssm\GeV, the
  superstring-inspired $\mathrm{Z}^\prime_\psi$ below \limitZpsi\GeV,
  and, for values of the coupling parameter $k/\overline{M}_{\rm Pl}$
  of 0.05 (0.1), Kaluza--Klein gravitons below \limitGlow\
  (\limitGhigh)\GeV.

\end{abstract}

\maketitle
\thispagestyle{fancy}


\section{\label{introduction}INTRODUCTION}

Many models of new physics predict the existence of narrow
resonances, possibly at the TeV mass scale, that decay to a pair of
charged leptons. 
We present results of a search for resonant signal
that can be detected by the Compact Muon Solenoid (CMS)~\cite{cms} detector at
the Large Hadron Collider (LHC)~\cite{lhc} at CERN. 
The results were obtained from an analysis of
data recorded in 2011, corresponding to an integrated luminosity of \anaLumi,
obtained from $\Pp\Pp$ collisions at a centre-of-mass energy
of 7\TeV.  The complete analysis is reported in \cite{pas}.

We perform a generic shape-based search for a narrow resonance, making no 
assumption on the absolute background rate.
The Sequential Standard Model $Z'_{SSM}$ with standard-model-like couplings, the
$Z'_\psi$ predicted by grand unified theories~\cite{zprime},
and Kaluza--Klein graviton excitations arising in 
the Randall--Sundrum (RS) model of extra
dimensions~\cite{rs} were used as benchmarks.

\section{LEPTON SELECTION}

The reconstruction, identification, and calibration of muons
and electrons
follow standard CMS methods.
%~\cite{MUO-10-004-PAS,EWK-10-002-PAS}. 
Combinations of test beam, cosmic-ray muons, and 
data from proton collisions 
have been used to calibrate the relevant detector systems
for both muons and electrons.

For both the dimuon and dielectron final states, 
two isolated same-flavour leptons that pass
the lepton identification criteria are required.
The two charges are
required to have opposite sign in the case of dimuons (for which a charge
misassignment implies a large momentum measurement error), but not in the
case of dielectrons (for which charge assignment is decoupled from
the electromagnetic calorimeter (ECAL) energy measurement). 
An opposite-charge requirement for
dielectrons would lead to a loss of signal efficiency of a few percent.
The electron sample requires at least one electron candidate in the ECAL
barrel (pseudorapidity range $|\eta|<1.479$), because events with both electrons in the endcaps ($1.479<|\eta|<2.5$) have a
lower signal-to-background ratio as a result of a higher rate of jets faking electrons.
For the muon sample muons reconstructed in full muon system coverage 
region ($|\eta|<2.4$) are used in the analysis.

The performance of the detector systems is 
established using measurements of standard model (SM) $\PW$ and $\cPZ$ processes 
with leptonic final states
%~\cite{EWK-10-002-PAS} 
and using traversing cosmic-ray muons.
%~\cite{MUO-10-004-PAS}.
The dimuon mass resolution is estimated to be 
4\% at 500\GeV ~and 7\% at 1\TeV. 
The dielectron mass resolution is more or less flat above 500 GeV. 
At 1 TeV the current mass resolution when
both electrons are in the barrel acceptance is 1.3\%, and when one
electron is in the barrel and the other within the endcap acceptance
the resolution is 2.4\%. 

\section{BACKGROUNDS}

The most prominent SM process that contributes to the dimuon and
dielectron invariant mass spectra is Drell--Yan production
($\cPZ{/}\gamma^*$); there are also contributions from $\ttbar$,
$\tq\PW$, diboson,
and $Z\rightarrow\tau\tau$ processes. In addition, jets 
%and non-prompt leptons from hadronic decays
may be misidentified as leptons and contribute to the dilepton
invariant mass spectrum through multi-jet and vector boson + jet final
states. The possible contamination from diphotons faking dielectrons has been 
found to be negligible. 

\begin{figure*}[t]
\centering
\includegraphics[width=85mm]{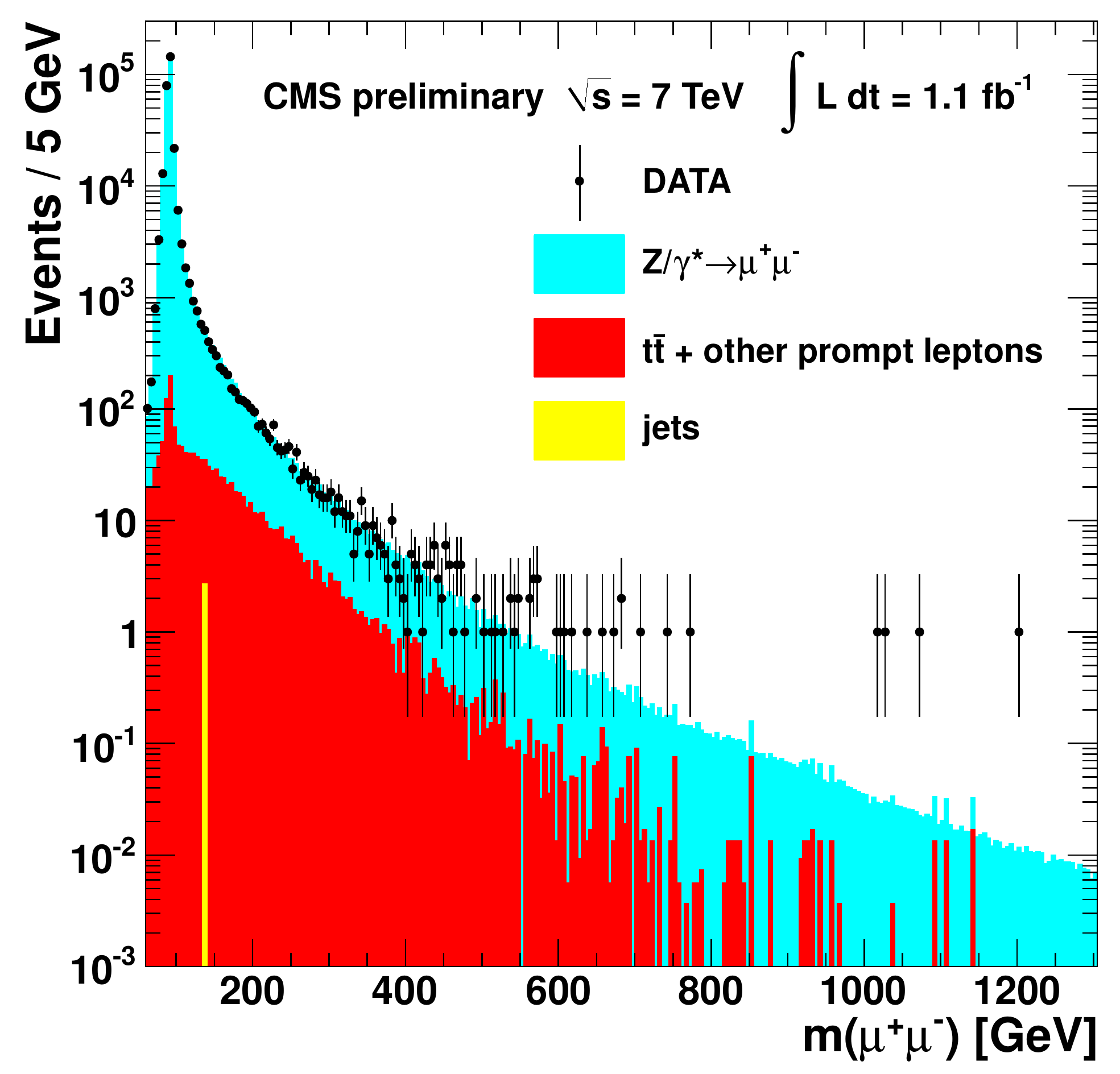}%
\includegraphics[width=85mm]{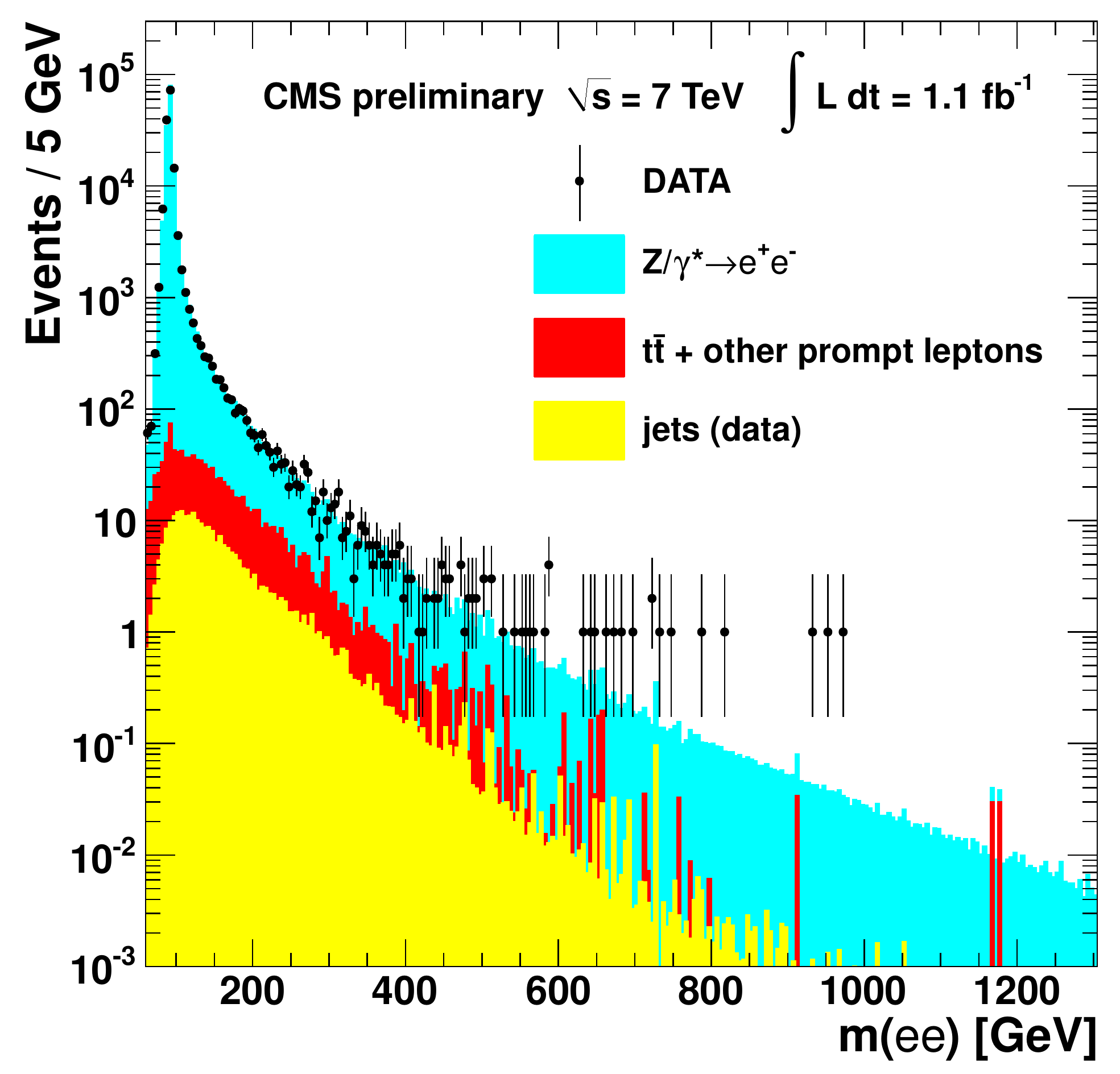}
\caption{\label{fig:spectra}
Invariant mass spectrum of $\Pgmp\Pgmm$ (left) and $\Pe\Pe$ (right) events. The points with
error bars represent the data. 
The uncertainties on the data points (statistical only) 
represent 68\% confidence intervals for the Poisson means. 
The filled histograms represent the expectations from
SM processes: $\cPZ{/}\gamma^*$, $\ttbar$, other sources of 
prompt leptons ($\tq\PW$, diboson production, $\cPZ\to\Pgt\Pgt$), and
the multi-jet backgrounds. 
}
\end{figure*}

To estimate contribution from Drell--Yan production 
the shape of the dilepton invariant mass spectrum is obtained from
%Drell--Yan production using a 
Monte Carlo (MC) simulation based on the
PYTHIA \cite{pythia} 
event generator.   
The simulated invariant mass spectrum is normalized to the data using
the number of events in the mass interval of 60--120\GeV.

The dominant non-Drell--Yan electroweak contribution to high
$m_{\ell\ell}$ masses is {$\ttbar$}; in addition there are contributions
from $\tq\PW$ and diboson production. Below the $\cPZ$ peak,
$\cPZ \rightarrow \Pgt\Pgt$ decays also contribute.
Here we refer to leptons coming from the decay of a $\PW$ or a $\cPZ$ as prompt leptons.
All these processes are flavour symmetric, producing twice as
many $\Pe\Pgm$ pairs as $\Pe\Pe$ or $\Pgm\Pgm$ pairs.
The invariant mass spectrum from
$\Pe^\pm\Pgm^\mp$ events is expected to have the same shape as that of same
flavour $\ell^+\ell^-$ events but without significant contamination
from Drell--Yan production.

A good agreement was established between
the observed and predicted $\Pe^\pm\Pgm^\mp$ distributions, which provides a validation of the
estimated contributions from the high-mass prompt
lepton backgrounds obtained using MC simulation.

A further source of background arises when objects are falsely
identified as leptons.  The misidentification of jets as
leptons, the principle source of such backgrounds, is more likely to
occur for electrons than for muons.

Backgrounds arising from jets that are misidentified as electrons
include $\PW\to \Pe\cPgn$ + jet events with
one jet misidentified as an electron, and also multi-jet events with
two jets misidentified as electrons.
Contribution from this background to the dielectron spectrum
is estimated from the data 
based on the rate at which a jet can be misidentified 
as an electron.

Contribution from misidentified muons to the dimuon spectrum is estimated from MC, 
validated using similar method as for electrons, and in addition validated 
by analyzing same-charge muon pairs mass spectrum.

The $\Pgmp\Pgmm$ data sample is susceptible to contamination from
traversing cosmic-ray muons, which may  be misreconstructed as a pair
of oppositely charged, high-momentum muons.
Cosmic-ray events can be removed from the data sample
due to their distinct topology 
(collinearity of two tracks
associated with the same muon),
and their uniform distribution of impact parameters with respect to the collision vertex.

\section{DILEPTON INVARIANT MASS SPECTRA AND RESULTS}

The measured dimuon and dielectron invariant mass spectra are displayed in
Figs.~\ref{fig:spectra}(left) and (right) respectively.
The expectations from the various background sources,
$\cPZ{/}\gamma^*$, $\ttbar$, other sources of  prompt leptons ($\tq\PW$, diboson
production, $\cPZ\to\Pgt\Pgt$) and multi-jet events
are overlaid. 
The prediction for Drell--Yan production of $\cPZ{/}\gamma^*$
is normalized to the observed $\cPZ\to\ell\ell$ signal.  All other MC
predictions are normalized to the expected cross sections.

\begin{figure}[t]
\includegraphics[width=78mm]{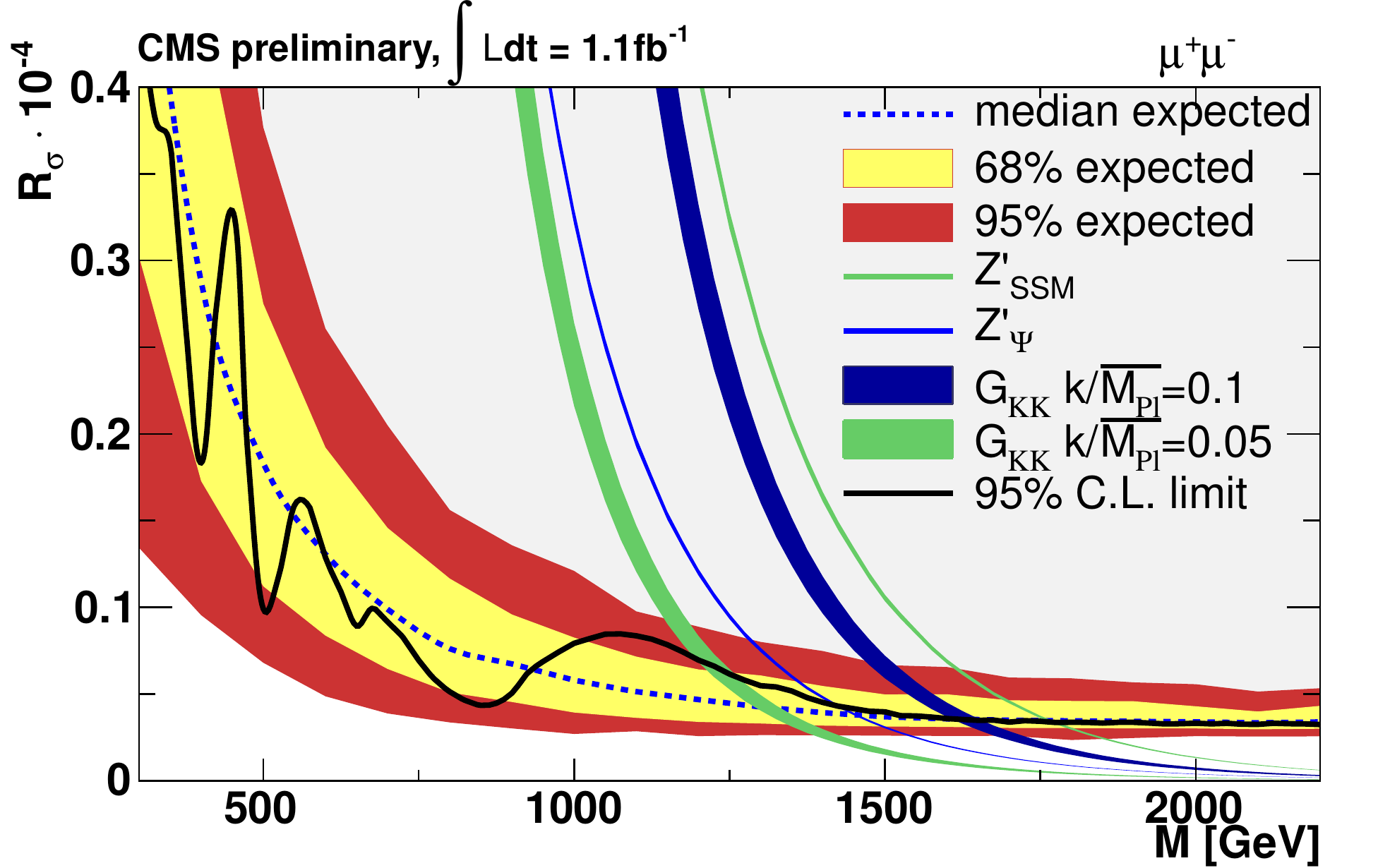}
\includegraphics[width=78mm]{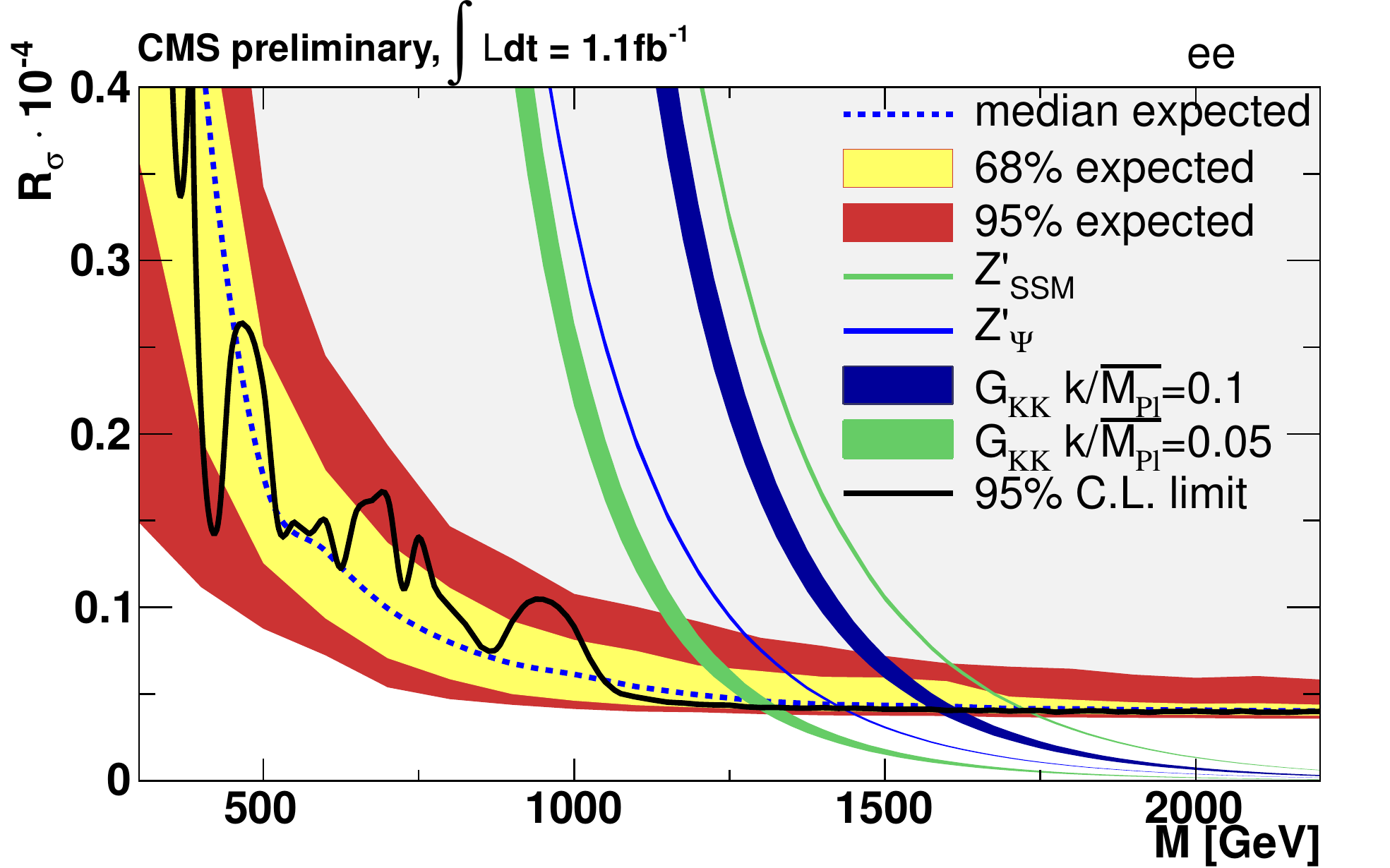}
\includegraphics[width=78mm]{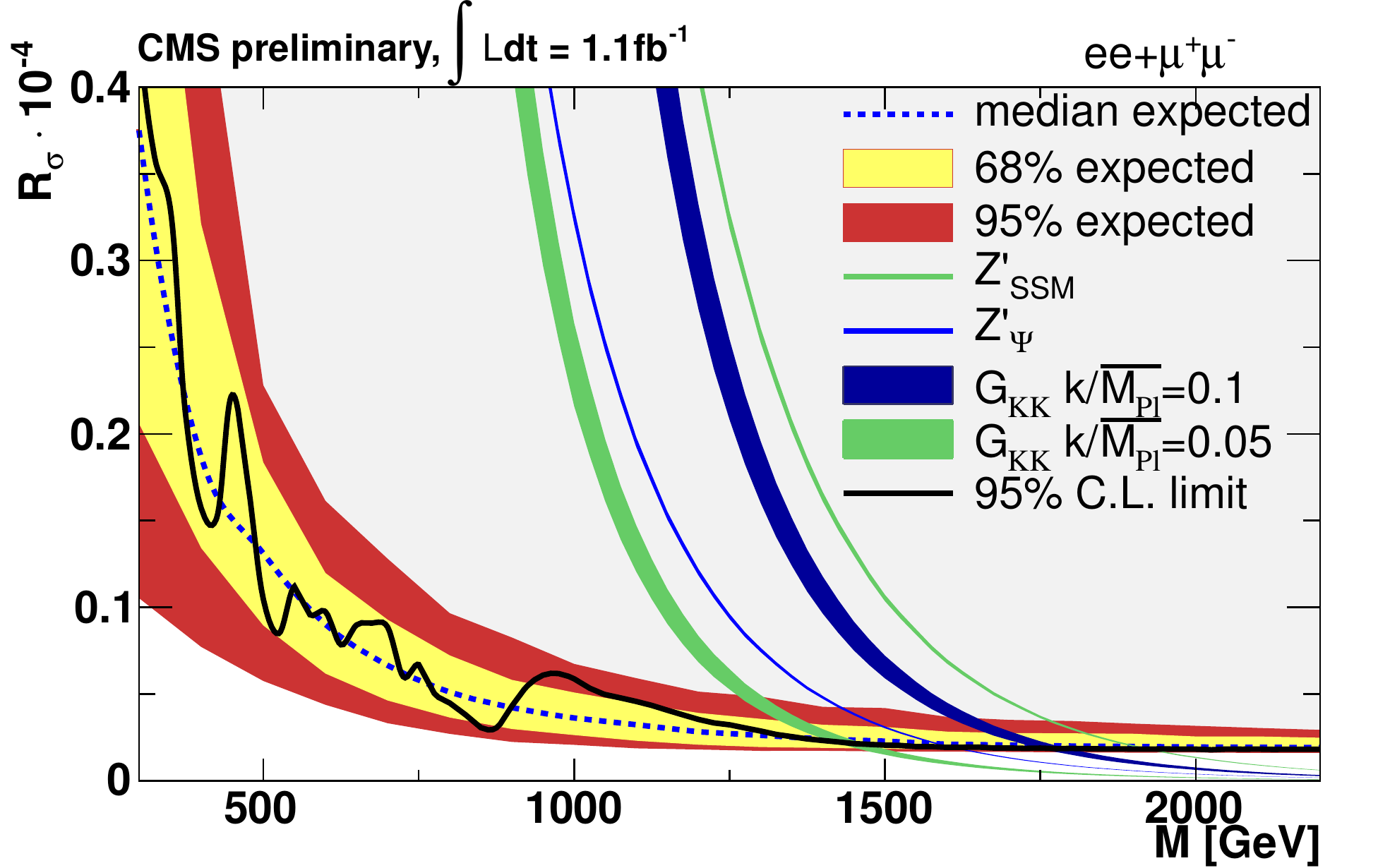}
\caption{\label{fig:limits}
 Upper limits as a function
of resonance mass $M$,
on the production ratio $R_{\sigma}$ of
 cross section times branching fraction into lepton pairs
  for $\ZPSSM$ and $\GKK$ production and $\ZPPSI$
 boson production. The limits are shown from (top) the $\Pgmp\Pgmm$ final
 state, (middle) the $\Pe\Pe$ final state and (bottom) the combined dilepton result.
Shaded yellow and red  bands correspond to the $68\%$ and  $95\%$ quantiles 
for the expected  limits.
The predicted cross section ratios are shown as bands, with widths
indicating the theoretical uncertainties. 
}
\end{figure}

Good agreement is observed between data and the
expectation from SM processes over the mass region above the Z peak,
and limits are set on the possible
contributions from a narrow heavy resonance.
The parameter of interest is the ratio of the products of cross sections
and branching fractions:
\begin{equation}
\label{eq:rsigma}
R_\sigma = \frac{\sigma(\Pp\Pp\to \cPZpr+X\to\ell\ell+X)}
                {\sigma(\Pp\Pp\to \cPZ+X       \to\ell\ell+X)}.
\end{equation}

By focusing on the ratio
$R_\sigma$, we eliminate the uncertainty in the integrated
luminosity, reduce the dependence on experimental
acceptance, trigger, and offline efficiencies, and generally obtain a more robust result.

Confidence intervals
are computed using 
both frequentist 
and Bayesian approaches.
The upper limits on $R_{\sigma}$ (Eq.~\ref{eq:rsigma}) from the various approaches
are similar, and we report the Bayesian result 
for definiteness.
From the dimuon and dielectron data, we obtain the upper limits on the cross section
ratio $R_{\sigma}$ at 95\% confidence level (C.L.) shown in
Figs.~\ref{fig:limits}(upper) and (middle), respectively, and the combined
limit (bottom).

\section{CONCLUSION}

The CMS Collaboration has searched for narrow resonances in the
invariant mass spectrum of dimuon and dielectron final states in event
samples corresponding to an integrated luminosity of \anaLumi,
taken at a centre-of-mass energy of
$7$~TeV.  The spectra are consistent with expectations from the standard model
and upper limits have been set on the cross section times branching fraction for
$\cPZpr$ into lepton pairs relative to standard model $\cPZ$ boson
production. Mass limits have been set on neutral gauge
bosons $\cPZpr$ and RS Kaluza--Klein gravitons $\GKK$.  A $\cPZpr$
with standard-model-like couplings can be excluded below
\limitZssm\GeV, the superstring-inspired $\ZPPSI$ below
\limitZpsi\GeV, and RS Kaluza--Klein gravitons below \limitGlow\
(\limitGhigh)\GeV ~for couplings of 0.05 (0.10), all at 95\%~C.L.

%% Create the reference section using BibTeX:

\end{document}